# Designing a Bit-Based Model to Accelerate Query Processing Over Encrypted Databases in Cloud

[1] Sultan Almakdi and [2] Brajendra Panda

[1,2] University of Arkansas, Fayetteville, AR 72701, USA
[1] Najran University, Najran, Saudi Arabia
[1,2]`{saalmakd, bpanda}@uark.edu`
[1]`smalmukadi@nu.edu.sa`

**Abstract.** Database users have started moving toward the use of cloud computing as a service because it provides computation and storage needs at affordable prices. However, for most of the users, the concern of privacy plays a major role as they cannot control data access once their data are outsourced, especially if the cloud provider is curious about their data. Data encryption is an effective way to solve privacy concerns, but executing queries over encrypted data is a problem that needs attention. In this research, we introduce a bit-based model to execute different relational algebra operators over encrypted databases at the cloud without decrypting the data. To encrypt data, we use the randomized encryption algorithm (AES-CBC) to provide the maximum-security level. The idea is based on classifying attributes as sensitive and non-sensitive, where only sensitive attributes are encrypted. For each sensitive attribute, the table's owner predefines the possible partition domains on which the tuples will be encoded into bit vectors before the encryption. We store the bit vectors in an additional column in the encrypted table in the cloud. We use those bits to retrieve only part of encrypted records that are candidates for a specific query. We implemented and evaluated our model and found that the proposed model is practical and success to minimize the range of the retrieved encrypted records to less than 30% of the whole set of encrypted records in a table.

## 1 Introduction

Nowadays, cloud computing is an attractive computation environment for both individuals and organizations since it provides a scalable data storage and high-performance computing unites. These features were not affordable for most individuals and small companies; so previously, only big companies could own such unites. With the presence of cloud computing, this problem has been solved as users can rent storage and computation unites as needed at an affordable price. Moreover, the majority of cloud providers offer database as a service, in which users and companies outsource their data and can access them anytime, from anywhere. However, people have expressed concern about their privacy when outsourcing sensitive data, as privacy breaches are one of the most common threats in the cloud-computing environment. For example, untrustworthy cloud providers can steal personal customer information such as emails, addresses, and phone numbers and sell them to third parties. Thus, users receive annoying advertisements through emails, the mail, and their phones. Furthermore, if an attack targets a cloud provider, the attackers can gain access to customers' sensitive personal information such as their social security numbers (SSNs). This has serious consequences, as the criminals can impersonate customers in different situations such as financial transactions like phone banking. As a result, sensitive data are restricted from being processed or sold to a third party. Therefore, the cloud-computing environment could become unattractive for consumers if there are no available appropriate solutions for privacy breach and security issues. This issue must be addressed if cloud providers are to gain the trust of users and organizations so that they will outsource their sensitive data without worrying about data leakages.

To solve the problem of privacy breach, data encryption is the only way to ensure that cloud providers cannot learn from the data they store. Different researchers have proposed various models for user-side data encryption, wherein data encryption happens before outsourcing, and decryption happens on the user's side. The problem with this technique is that it conflicts with critical functionalities of cloud environments (e.g., searching for a numeric range of data). Other researchers use what is called onion layer encryption where each data item is encrypted with more than one encryption algorithm to support various query types [24]. However, in case of huge data sets, the penalty is the computational burden, as each data item might have to be decrypted more than once.

In this proposed research, we design a model to execute different relational algebra queries over encrypted data. The proposed model deals with both encrypted numeric data and textual data. We introduce the query manager (QM) component—light software for single users and a trusted server in organizations—which performs the encryption, decryption, and queries translation, leaving a minimal amount of work for users. In addition, in our model, we split the computation into two sides: a cloud provider side and a client side, where we move the majority of the computation to the cloud provider by translating queries into appropriate ways to deal with encrypted data without decrypting them. We design an algorithm for each query category (e.g., select, join, union, intersection, etc.) to enable cloud providers to execute such query categories over encrypted tuples. Further, we encrypt sensitive data in tables with the randomized Advanced Encryption Standard (AES-CBS) encryption algorithm that is neither deterministic (each plaintext always has the same cipher text) nor order preserving in order to provide the highest level of security and privacy.

We classify attributes to sensitive and non-sensitive. Sensitive attributes can be partitioned into partition domains (PDs) by the owner of the table based on possible values for the attribute. The query manager (QM) creates a data structure for the partitions domains (PDs), and then creates a bit vector whose length is equal to the number of PDs and maps each PD to a bit position. The encrypted table in the cloud will have an additional column to store those bits for each tuple. We use those bits to retrieve only candidate encrypted tuples of a query from the cloud. Also, we make our model resistances to different attacks, such as inference attack, that could happen at the cloud site where the attacker could exploit the presence of the bits to infer the possible data distributions. The rest of this paper is structured as follows: in section 2, we discuss related work and previous models. Then, in section 3, we explain the model in detail. We describe how we implemented and evaluated our model in section 4, and we provide the conclusions of this work in section 5.

## 2   Literature Review

CryptDB is a system proposed by Popa et al. [23] as the first practical system for executing different Standard Query Language (SQL) queries over the encrypted databases. The model was designed to resist two different possible attacks, cloud attacks, and proxy attacks. They introduce onion layers encryption that is encrypting each data item by more than one encryption algorithm to support multiple SQL queries. Moreover, they present new algorithms, including one to handle the join operation. In CryptDB, the primary purpose of the proxy is to perform the crypto processes on behave of the users. One of the downsides of CryptDB is the high computation burden because each data item must be encrypted and decrypted more than once.

Liu et al. in [17] propose the FHOPE system to support complex SQL queries over encrypted data while resisting homomorphic order-preserving attacks. It allows cloud providers to run arithmetic and comparison operators over encrypted data without repeating the encryption. The limitation of this work is that the authors conducted their experiments based on tables with less than 9,000 records. In order to show the efficiency and scalability of the system, using tables with larger numbers of records (e.g., 100,000; 500,000; and 1,000,000) would better in terms of assisting the efficiency and overhead since this requires more decryption processes. A variety of studies related to this system can be found at [7,8,16,18,19,24,30].

Cui et al. propose P-McDb [3], a privacy-preserving search system that enables users to run queries over encrypted data. It uses two clouds to prevent inference attacks, one for data storing and searching and one for database re-randomizing and shuffling. It supports partial searches of encrypted records (as opposed to total searches), a feature referred to as a sub-linear manner. P-McDb can be used by multiple users, and in the case of a user revocation, the data cannot be re-encrypted. In this system, the communication between the two clouds could introduce more delays when compared with other systems, like [23]. Other proposed models in this matter are described in [2,4,5,6,13,32].

In [10], the authors have created a system that executes relational algebra operators over encrypted data. They add identifiers for attributes to enable service providers to execute queries over data. They split the computation between the end user and the service provider, with the majority of the computation moved to the service provider's side. The limitation of this work is that each tuple is encrypted and stored as a string at the service provider's site. This prevents some of the relational algebra operators (e.g., projection) from being executed on the provider's side, and in some cases, whole encrypted tuples must be returned and decrypted (see [9,14,15,22] for related research).

Osama et al., in [21], propose different approaches for partitioning attributes of tables into multiple domains. They tested their approaches and found that they introduced different delays. Their approaches are based on an order-preserving mapping function, which enables cloud servers to execute different operators as equality operators. The limitation of this work is that they did not consider textual data in their experiments, which would prove that their approaches are practical for relational databases. They should also have considered relational algebra operators.

In [27], the authors proposed SDB that is a model based on dividing data to sensitive and non-sensitive data; only sensitive data are encrypted. The idea is to split the sensitive data into two shares, one share is kept by the data owner (DO) and the second is kept by the service provider (SP) assuming the SP is untrusted. In SDB, the SP cannot reveal anything from the share that it has. Besides, the SDB allows different operators to share the same encryption, providing secure query processing with data interoperability. Similar work can be found in [20, 31, 28].

Bucketization is a method that requires the indexing and partitioning of the encrypted data into more than one bucket. Each bucket has an identification (ID) and holds a set of encrypted records ranging from the minimum to maximum value. The index can be used to execute SQL-style queries over the encrypted data [26]. Various models have been done based on this approach [11, 12, 25, 29].

## 3  The Model

The main goal of this research is to design a model that protects data in cloud servers from being accessed by curious cloud providers or malicious attackers. Each database table in this model will be encrypted before being outsourced to the cloud, using the randomized encryption algorithm, AES-CBC; concurrently, we enable the cloud server to execute queries over the encrypted data. Unlike traditional database encryption models where the user is the one who encrypts and decrypts the data, as in our previous work [1], our model uses a query manager (QM) as an intermediary between users and the cloud. As shown in Fig. 1, our model features the QM as a light software for single users or a trusted server that resides in the private cloud, for users within organizations. We design the QM to perform the computations, creating minimal work for the user(s). In the following sub-sections, we explain the model in detail.

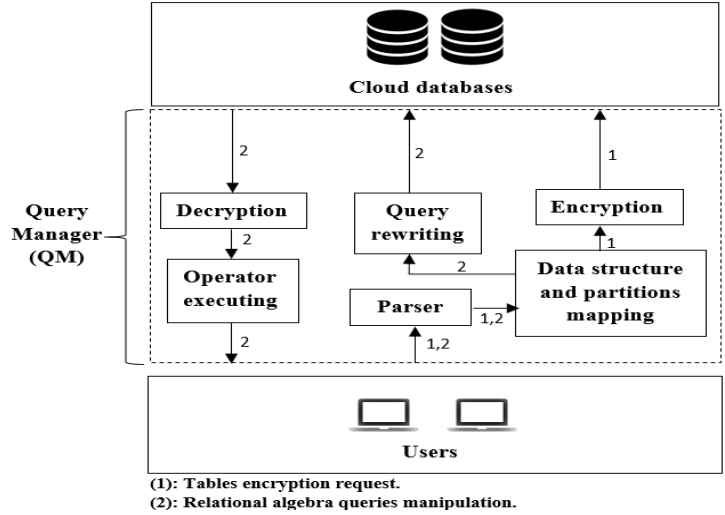

**Fig. 1.** System architecture

### 3.1 The Details of the Proposed Model

We divide the attributes into sensitive and non-sensitive attributes. The sensitive attributes are the only attributes that are encrypted; because if they are not, they would leak private information, whereas non-sensitive attributes do not leak private information. We classified sensitive attributes, where data must be encrypted, into two types: attributes that have limited distinct values, such as student_rank, and attributes that may have too many distinct values (ranges), such as salary. Therefore, each attribute is partitioned into multiple partitions, and the table's owner is the one who predefines these partitions before beginning the encryption process. Then the QM builds a data structure of all partitions for each table, as in Fig. 2. After that, the QM creates a bit vector (BV) for each record where its length equals the total number of partitions for all the sensitive attributes. Then we store those bits in an additional attribute(s) in the encrypted table at the cloud, unlike our previous work [1] where we store bit vector(s) in the QM. We use the bits to retrieve only candidate encrypted tuples for a query and element unrelated tuples retrieval. The QM performs the following steps for creating bit vectors (BVs) and encrypting tuples: 1) The table's owner defines partitions domains for each sensitive column and sends a creation request to the QM. 2) According to the PDs, the QM creates a data structure for the table to easily translate queries in the future. 3) The QM takes each record $j$ and creates a bit vector $BV_j$ whose length is equal to the total number of partition domains for all sensitive columns for a table T, map each bit's position to a single domain partition, and initializes all bits with 0's. 4) A bit $i$ in $BV_j$ is set if and only if the data item $d$ under column $A$ equals or belongs to partition domain $A_i$. 5) The QM shall encrypt sensitive columns' data by AES-CBC and send encrypted data to the

66

Table 1. Students Table

| ID  | Name  | Rank      | Visa type | Department           |
|-----|-------|-----------|-----------|----------------------|
| 110 | Alice | freshman  | F1        | Computer science     |
| 111 | Sara  | senior    | J1        | Computer engineering |
| 112 | John  | junior    | None      | Information system   |
| 113 | Ryan  | Sophomore | J2        | Math                 |

cloud database, along with the $BV_j$ from the previous step. 6) The steps above are repeated for all records in Table T. In the cloud, the encrypted table has an additional column (we call it the reference column) to keep a BV for each record. The data type of the reference column that stores BVs is a BIT(n), where n is the number of bits. So, we don't worry about the size growth of the BVs since they are in bits and stored at the cloud. Do the fact that bits attributes can hold up to 64 bits, we could have more than one attribute to store bits. Then when rewriting the query, the QM can take the bits position find the reminder of 64 (e.g. when the bit position is 80, then 80%64 =16 which means the bit's position in the second bits attributes is 16). In this way, we can exploit the high computation speed and the massive capacity provided by the cloud to process and store BVs. The longer the BVs are, the less encrypted records are retrieved.

**Security of proposed model.** Although it's not an easy process for an adversary (either malicious attacker or curios cloud provider) to learn the distribution of data nor infer the possible values of a column from the BVs, we add another security step to address this vulnerability. The solution is based on encrypting the names of the tables and the sensitive columns by a deterministic encryption algorithm, namely AES-SIV, where the ciphertexts for each plain text are always identical. The reason for that is to maximize the security to the highest level by making it impossible for the adversary to infer or learn from the BVs column the possible values while we enable the cloud provider to execute our queries. For instance, the column "Students-Rank" have limited possible values like grad, senior, junior, etc. Then the adversary could infer those even though they are encrypted with AES-CBC (which always produces different ciphertexts for a plain text). Details and security analysis of AES-SIV can be found in [33]. Note that, having more than ciphertexts share the same prefix is a not concern since the adversary cannot get the plain text unless he obtains the SK (Secrete key), and the SK isn't passed to the cloud in our case. Therefore, by encrypting the names of tables and sensitive columns, we ensure that the adversary can learn nothing from the BVs that we store at the cloud.

**Example.** Suppose we want to encrypt the students table presented in Table 1: the QM encrypts the table name and the names of sensitive columns by AES-SIV using user's secret key (SK) and create a new table at the cloud. Then the QM takes the first record and parses it to get the values. The name is "Alice," where the first letter is "A," which falls under the range "A–F," so the bit in the position $2^{19} = 524288$ will be set to 1. Then, the rank is "freshman", that will set the bit that mapped to position $2^{15}$ to 1. Next, the visa type is "F1," so QM sets bit with position $2^{10}$ to 1; the same process occurs for the department column, which will set position $2^5$ to 1. Now, the bit vector (BV) is

| SCs | Name's first letter | | | | Rank | | | | | Visa Type | | | | | Department | | | | | |
|---|---|---|---|---|---|---|---|---|---|---|---|---|---|---|---|---|---|---|---|---|
| PDs | A-F | G-L | M-R | S-Z | Fresh | Sen | Jun | Soph | Grad | F1 | F2 | J1 | J2 | None | CS | CE | IS | Bus | Math | other |
| BPs | 524288 | 262144 | 131072 | 65536 | 32768 | 16384 | 8192 | 4096 | 2048 | 1024 | 512 | 256 | 128 | 64 | 32 | 16 | 8 | 4 | 2 | 1 |
| | $2^{19}$ | $2^{18}$ | $2^{17}$ | $2^{16}$ | $2^{15}$ | $2^{14}$ | $2^{13}$ | $2^{12}$ | $2^{11}$ | $2^{10}$ | $2^{9}$ | $2^{8}$ | $2^{7}$ | $2^{6}$ | $2^{5}$ | $2^{4}$ | $2^{3}$ | $2^{2}$ | $2^{1}$ | $2^{0}$ |

**SCs:** denotes sensitive columns.
**PDs:** denotes partitions domains.
**BPs:** denotes bits' positions (e.g. the 1st bit in the bit vector is the bit that is located at position 1 while the last bit in the bit vector is located at position 524288.

**Fig. 2.** Data structure and mapping information at the QM for student table.

"10001000010000100000". The QM will do this for all the records in the table. The encrypted table in the cloud will look like Table 2. To retrieve encrypted data from the encrypted table, when a user submits a query to the QM as

"*select* name from student where department = "computer science"

The QM processes the query as in follow:
1) The QM encrypt the table's name and the sensitive columns' names using the user's SK to obtain the cipher texts as below:
    $CT_1 = E_{AES-SIV}$ ("Students", SK)
    $CT_2 = E_{AES-SIV}$ ("Name", SK)
2) The QM looks up the data structure, finds the bits' representations, then rewrites the query as "*select* $CT_2$ from $CT_1$ where reference&32 > 0"
3) The cloud will return only encrypted tuples that have the bit at position 32 ($2^5$) = 1.

### 3.2 Condition Rewriting

To enable the cloud database server (CDBS) to search over encrypted data, the QM must translate the query conditions in ways that the CDBS can implement them over encrypted tuples. This is a very important step before implementing queries of any query category, such as joins, union, and aggregations. The QM translates queries based on the data structures (DSs) of the table(s) involved in the query. As in [9], we have three kinds of conditions:

1. A condition that has a column and a value, for example,
    Name = "Alice".
2. A condition that has columns only, such as Name = Name.
3. A condition that involves more than one condition, such as Name= "Alice" OR visa type = "F1".

In the first and second kinds, there are five possible operations $\{>, <, =, \geq, \leq \}$. The operation in the third kind is limited to $\{\vee, \neg, \wedge\}$. In the first kind, column and value, the QM parses the query to extract the table's name, columns, and values. Then, for each value $v$, the QM finds the partition domain $PD_i$ that value $v$ equals or belongs to, and keeps its position. If the operation is "=", then it is straightforward, and the QM substitutes values with its corresponding bit position. We explain this case in detail under selection operation in the next sub-section. If the operation is "> or ≥", then the QM finds the positions of all PDs that are less than $v$. Therefore, the QM translates the query by substituting value $v$ with the positions separated by OR and "> or ≥" with "=". For instance, suppose we have an employee table, and it has a salary column that is partitioned into three PDs {(1000-5000), (5001-9000), (9001-12000)}. Assume that the bits' positions are 32, 16, and 8, respectively. Suppose that the query is to find "salary > 8000". Now the query condition is translated into "Reference &16>0 OR Reference &8>0". Note that the PD that $v$ is belongs to must be retrieved too with the PDs that must to be retrieved from the cloud, because the PDs are ranges. For the rest of the operation "< or ≤", we treat them in a similar way as "> or ≥", but by taking PDs that are less than or equal to $v$.

In the second kind, the condition that contains only columns, the QM finds all PDs' positions for every table involved in the condition. If the operation is " = ", the QM rewrites the query condition by substituting the columns' names with pairs ($PD_i$ from column A ∧ $PD_i$ from column B, where $PD_i=PD_j$) of the DPs' positions separated by the OR operation (pair1 V pair2 V pair3 …). To illustrate this, consider Table 1 (students table) and assume that we have another table (students_info table) and both tables have the same data structure as in Fig.2. Suppose the condition is "where students.name = students_info.name". Let's say that,

$CT_1$ = $E_{AES-SIV}$ ("Students", SK)

$CT_2$ = $E_{AES-SIV}$ ("students_info", SK)

We assume that the owner of the two tables is one who owns the *SK*. The translation will look like the following:

"where (($CT_1$.Reference&524288>0 AND $CT_2$.Reference&262144>0) OR
($CT_1$.Reference&262144>0 AND $CT_2$.Reference&131072>0) OR
($CT_1$.Reference&131072>0 AND $CT_2$.Reference &65536>0)  OR
($CT_1$.Reference&65536>0 AND $CT_2$.Reference &32768>0))"

We do the same if the columns are numeric, and the operation sign is {>, <, ≥, ≤ }, where we arrange the pairs in a way to satisfy the order condition.

In the third kind, the condition that contains more than one condition, each condition can be one of the above two kinds separated by AND or OR operations. Therefore, once all conditions are translated using the two translation methods, we need to combine the translation result with the AND or OR operation as in the query. For example, consider Table 1. Suppose the condition is (where Name = "Alice" OR Visa type = "J2"). This condition has two conditions with the OR operation, so the translation can be (where reference&524288>0 OR reference&128 >0). We perform the same method with the AND operation.

**Table 2.** $E_{AES\text{-}SIV}$ ("Students", SK), encrypted students table at the cloud. Note that the names of the columns are encrypted too using AES-SIV where $ct_n = E_{AES\text{-}SIV}$ ("Name", SK), $ct_r = E_{AES\text{-}SIV}$ ("Rank", SK), $ct_{vt} = E_{AES\text{-}SIV}$ ("Visa type", SK), $ct_d = E_{AES\text{-}SIV}$ ("Department", SK).

| ID  | ct$_n$ | ct$_r$ | ct$_{vt}$ | ct$_d$ | Reference          |
|-----|--------|--------|-----------|--------|--------------------|
| 110 | *&^    | *_^%   | */d       | ^%^H   | 10001000010000100000 |
| 111 | %^&    | /+$    | &^/       | &&%$   | 00010100000100010000 |
| 112 | )(#    | %$/*   | +-*&      | )*#R   | 01000010000001001000 |
| 113 | $#!    | !@~K   | */f       | @$%*   | 00100001000010000010 |

### 3.3 Relational Algebra Operators

In this subsection, we describe the set of relational algebra operators the proposed model can support. Most previous studies focused on the select operation, and only a few considered other relation algebra operators, like aggregation, union, intersection, difference, sort, and duplicate elimination. This is because it is challenging to facilitate a model that can support all relational algebra over encrypted data. Therefore, to make the proposed model practical, we provide an algorithm to execute each operator, such that as much of the computation as possible is moved to the cloud provider, leaving minimal work for the QM. The proposed algorithms translate the queries in a way that they filter out only unrelated encrypted records.

**Select.** The easiest way to implement the *select* operation over encrypted data is to retrieve the whole encrypted table from the cloud, decrypt it at the QM, and then execute the select operation on the data after they have been decrypted. However, this method is not practical in the case of huge tables, because it adds more computation burden on the QM. In an alternative method, in the proposed model, we move as much computation as possible to the cloud database server (CDBS) which has efficient computation unites, so we aim to minimize the range of retrieved encrypted records as much as possible. This can be done as follows: The QM gets the user's query and rewrites it according to tables' data structures (DSs), and then sends the translated select query to the CDBS (note that the QM translates the clauses that involve only sensitive columns). Then the CDBS executes the translated query and sends back the result (encrypted tuples for all candidate records). Now, the QM decrypts and applies the selection operation to them to filter unrelated data before returning the result to the user. We illustrated the process in the example in the previous sub-section (3.1). Algorithm 1 demonstrates the steps that are performed by the QM to execute the select operator.

**Join.** To enable the proposed model to support the *join* operator, we must consider deferent cases for the join condition: 1) The join condition involves only non-sensitive columns, 2) the join condition involves only sensitive columns that have limited distinct values, and 3) the join condition involves sensitive columns that may have too many distinct values. The first case is straightforward, as the QM is required only to forward the query to the cloud database server (CDBS). Then it decrypts the result and removes the duplication.

| | **Algorithm 1: Select operator** |
|---|---|
| 1: | Input < Table name, List of all columns, List of all data items $di$(s) > |
| 2: | **If** none of columns $c_i$ mentioned in the query are sensitive |
| 3: |    **Forward** the query to CDBS |
| 4: |       **Decrypt** encrypted data |
| |       **Send** result back to the user |
| 5: | **If** all column(s) $c_i$ mentioned in the query are sensitive |
| 6: |    **For** each data item $di$ / value $v$ being searched for under column $c_i$ |
| 7: |       **Find** the bit's position that mapped to the partition domain $PD$ that $di$ falls under in the table's data structure $DS_T$. |
| 8: |    **Rewrite** the query and substitute values by bits' positions. |
| 9: |    **Send** the translated query to CDBS to retrieve candidate tuples (CTs) |
| 10: |    **Decrypt** CTs then **implement** select again to filter out unrelated records. |
| 11: |    **Send** result back to the user |

In the second case, the QM uses the data structure DSs of the tables mentioned in the *join* condition to match the PDs of the tables. In Table $T_1$, each $PD_j$ under a sensitive column $SC_i$ is joined with each $PD_j$ under a sensitive column $SC_i$ in Table $T_2$ if and only if the $PD_j$ from $T_1$ equals the $PD_j$ from $T_2$. Note that the QM does this only for columns that are involved only in the join condition.

The third case is based on range PDs. The QM joins the PDs from both tables, and it ensures that each $PD_i$ of the SC from Table $T_1$ is joined with each $PD_i$ of the SC from Table $T_2$ if $DP_i$ has at least one element that is common. For example, suppose we have two tables, Table A and Table B. We want to join them by the salary column. Suppose the salary DPs in Table A are [$PD_1$ (10,000–15,000), $PD_2$ (15,001–20,000), $PD_3$ (20,001–25,000), $PD_4$ (25,001–30,000)], and for Table B are [$PD_1$ (10,000–20,000), $PD_2$ (20,001–30,000)]. The QM joins $PD_1$ from Table B with $PD_1$, $PD_2$ from Table A because $PD_1$ of Table B contains elements from both $PD_1$ and $PD_2$ of Table A, and so on. Then the QM rewrites the query and sends it to the cloud. The cloud returns the join result to the QM, which decrypts only the columns involved in the join condition, and checks whether the plain texts satisfy the join condition. Then the QM decrypts the whole tuple only if the two values satisfy the join condition. Otherwise, the QM skips to the next tuple. We do so to eliminate unnecessary decryption processes. The QM eliminates duplicates after finishing decrypting the whole result. Algorithm 2 shows the steps of the *join* operation.

**Aggregation.** In this operation, we have two cases: First, the condition clause is based on column(s) where its partitions domain PDs values are not ranges. In this case, the CDBS efficiently implements the aggregation operation over encrypted data, and sends back the result to the QM, which needs only to decrypt the result and send it back to the user. Note that the QM may not implement the aggregation operator again over the decrypted data, because the candidate tuples retrieved from the CDBS are the exact query result.

For example, consider Table 1, Table 2, and Fig.2; a query *select* count(*) from a student department where the department = "Computer science" is translated to "*select* count(*) from $E_{AES-SIV}$("Student", SK) where reference &32 >0", and it returns only the number of tuples that satisfy the condition, because the condition is based on the partition domain "computer science" that is not a range of values. The second case is the condition clause, which is based on column(s) where its partitions domains' values are ranges. In this case, in addition to the first case, the QM implements the aggregation operator again over decrypted candidate tuples. This is because the candidate tuples have tuples unrelated to the query, as the PDs are ranges. In the average operation, we leave the whole calculation to be done at the QM. However, we minimize the range of encrypted records retrieved from the CDBS by only those that satisfy the condition clause of the average query. Algorithm 3 below illustrates the steps.

| | **Algorithm 2: Join operator** |
|---|---|
| 1: | Input < Tables names, List of all columns, List of all data items *di*(s) > |
| 2: | **If** none of columns mentioned in the join condition are sensitive |
| 3: | **Forward** the query to CDBS |
| 4: | **Decrypt** encrypted data and **Remove** duplication |
| | **Send** result back to the user |
| 5: | **If** all columns mentioned in the query are sensitive |
| 6: | **If** the SCs have limited distinct values |
| 7: | **Join** the bit's position of each partition domain $PD_i$ of a $SC_k$ from table $T_m$ with the equivalent $PD_i$ of a $SC_x$ from table $T_n$ |
| 8: | **If** the SCs have range values |
| 9: | **Join** the bit's position of each partition domain $PD_i$ of a $SC_k$ from table $T_m$ with each $PD_i$ of a $SC_x$ from table $T_n$ if it has at least 1 common value. |
| 10: | **Rewrite** the query and substitute columns' names by bits' positions separated by OR operation. |
| 11: | **Send** the translated query to CDBS to retrieve candidate tuples (CTs) |
| 12: | **Decrypt** CTs and **Remove** duplication |
| 13: | **Send** result back to the user |

**Sorting.** Executing a sorting operator over encrypted data is not an easy process. However, in the proposed model, we can make the cloud server filter out unrelated records first according to the PDs of the data structure (DS). Then the QM decrypts the returned sorted result from the cloud and executes the sort operator over them again. The sorting computation needed from the QM before returning the final result to the user and after retrieving candidate tuples is significantly small, because the CDBS sends back only candidate records that fall under a certain partition domain(s). However, the CDBS will return a group of records that are not sorted except by partition domain. For example, to retrieve tuples that have income ranging from 50k to 60k, the cloud server returns all records that have income within this range unsorted, because they are encrypted. Therefore, the QM sorts them after the decryption process. Algorithm 4 shows the processes.

| | **Algorithm 3: Aggregation operator** |
|---|---|
| 1: | Input < Tables names, List of all columns, List of all data items $di$(s) > |
| 2: | **If** none of columns mentioned in the aggregation query are sensitive |
| 3: |     **Forward** the query to CDBS |
| 4: |         **Decrypt** encrypted data and **Remove** duplication |
| |         **Send** result back to the user |
| 5: | **If** all columns mentioned in the query are sensitive |
| 6: |     **Reconstruct** the query by substituting data items/ values by the bits' positions that mapped to the PDs that contain data items. |
| 7: |     **Send** the query to CDBS |
| 8: |         **If** the SCs in the aggregation query are not ranges |
| 9: |             **Decrypt** the result |
| 10: |             **Send** result back to the user |
| 11: |         **If** the SCs in the aggregation query are ranges |
| 12: |             **Decrypt** the result |
| 13: |             **Execute** aggregation operator again over the result |
| 14: |             **Send** result back to the user |

| | **Algorithm 4: Sort operator** |
|---|---|
| 1: | **Do** steps 1 to 14 from aggregation algorithm |
| | "Note that we need to substitute the term" aggregation" by "sort" |

**Duplicate Elimination.** In our model, to execute the duplicate elimination operator, we make the CDBS execute a selection query over encrypted data without duplication elimination *keyword* by just substituting the columns and values of the condition with the positions of the corresponding bits in the reference column. Then the QM selects *distinct* values before sending the results back to the user. For example, the query: select *distinct* Name from students where department = "Computer science" is translated to: *select* the name from $E_{AES-SIV}$ ("Student", SK) where reference &32>0. The CDBS returns all encrypted names where the bit position 32 is set to 1 in the reference column. Then the QM implements the DISTINCT or DISTINCTROW keyword over the decrypted records to eliminate the duplication. Algorithm 5 below shows the processes.

| | **Algorithm 5: Duplicate -Elimination operator** |
|---|---|
| 1: | **Do** steps 1 to 9 from select algorithm |
| 2: | **Decrypt** encrypted data |
| 3: |     **Execute** Duplicate -Elimination operator again over the result |
| 4: |     **Send** result back to the user |

**Project.** We can implement the project at the cloud database server (CDBS) by using the select operator. We cannot eliminate the duplication in the cloud; however, we can delay this step until the encrypted tuples arrive at the QM, which will be able to eliminate duplicates after decrypting the cipher texts. Note that we do not retrieve all the encrypted tuples' column values, as each column's value is encrypted separately. In this case, we illuminate unnecessary decryption processes. Algorithm 6 illustrates the processes.

| | **Algorithm 6: Project operator** |
|---|---|
| 1: | Input < Tables names, List of all columns, List of all data items $di$(s) > |
| 2: | **If** none of columns mentioned in the project query are sensitive |
| 3: |     **Forward** the query to CDBS |
| 4: |     **Send** result back to the user |
| 5: | **If** all columns mentioned in the project query are sensitive |
| 6: |     **Forward** the query to CDBS |
| 7: |     **Decrypt** the result |
| 8: |     **If** the term **"DISTINCT"** is present in the query |
| 9: |         **Implement** *distinct* over decrypted data |
| 10: |     **Send** result back to the user |

**Union.** Before explaining how the proposed model performs the union operation over encrypted data, we discuss the two fundamental conditions of the union operation. First, all tables involved in the union operation must have the same number of columns. Second, the domain of the $i^{th}$ column in Table A must be the same as the ith column in Table B. In the proposed model, the QM translates the union query, using the condition translation methods proposed in the previous section, to the CDBS. Then the CDBS implements the union operator, and it returns all tuples of both relations. Note that we cannot eliminate duplication in the cloud, because the PDs could have at least one PD as a range. For example, the first letter column for Name has four PDs {a-f, g-l, m-r, s-z}, and we cannot remove the duplication before the decryption process. Therefore, we defer the removal of duplicates until after the encrypted tuples are decrypted by the QM. We assume that all tables are encrypted with the same encryption key, because if each table were encrypted with a different key, that would result in incorrect decryption. In algorithm 7, we show the steps.

| | **Algorithm 7: Union operator** |
|---|---|
| 1: | Input < Tables names> |
| 2: | **Forward** the query to CDBS |
| 3: | **Decrypt** the result of union query |
| 4: |     **Implement** *distinct* over decrypted data |

**Intersection.** The intersection operation has the same two conditions as in the union operation. However, instead of retrieving all encrypted tuples of both tables (Table A and Table B), we retrieve only the tuples that are common to the tables. We must simulate the intersection operation, because it is impossible to apply it over encrypted tuples. To do that, in the CDBS, we use the inner join between the two tables, where the join condition is based on PDs that are not ranges, as range-based PDs return more candidate tuples. Then the CDBS executes a query to choose the tuples from Table A that exist in the join result. After that, the CDBS returns the joining result to the QM, which eliminates duplicates after the decryption process. For example, suppose that Table A and Table B have attributes (Name, Visa type, Rank). Then we perform the inner join operation where the join condition is "visa type = visa type AND Rank= Rank). Note that, in the condition, we substitute the columns' names with the positions of the corresponding bits to enable the CDBS to execute the operation. This filters out too many uncommon tuples in the cloud, which results in fewer decryption operations by the QM. Algorithm 8 below illustrates the processes.

| **Algorithm 8: Intersection operator** |
|---|
| 1: Input < Tables names> |
| 2: **Find** all common sensitive columns that are not ranges between the tables |
| 3: **For** each table $T_i$ |
| 4:   **Find** bits' positions of PDs |
| 5: **Initiates** an inner join query where the joining condition is based on the equivalence of the bits obtained from step 4 for each table. |
| 6: **Send** the query to CDBS |
| 7: **Decrypt** the result of join. |
| 8: **Implement** *distinct* over decrypted data to **Remove** duplication |
| 9: **Send** result back to the user |

## 4   Implementation and Evaluation

In this section, we describe how we implemented the proposed model and how we evaluated it. To implement the functions of the QM, we used Java to simulate each function in a class. The user submits a plain query to the QM, and the QM parses it, and rewrites it using predefined DSs. There is no need to modify the internals of the database, because the rewriting methods translate the users' queries in ways that can be executed by the CDBS. We used the MySQL server on the user's machine, and we used Java Database Connectivity (JDBC) as a connector from Java to MySQL. As a cloud, we created a MySql account at the university's server to serve as the cloud database server (CDBS).

To evaluate the efficiency of the proposed model, we conducted different experiments in which we measure the time from a plain query is submitted until the result is returned to the user. In some experiments, we submit a query and find the percent of retrieved encrypted records. To accurately measure the efficiency, we compared the proposed model with a traditional database system in which the data were in plaintexts.

In each experiment we performed, we ran queries on a table that held a certain number of records (10K and 20K,) where each table has six attributes (two of them are non-sensitive, two are sensitive but not ranges, and two are sensitive but ranges). That helped us to evaluate our model and determine how the number of records affected the delay time. We tried different queries, with each query coming from the following set of queries:

1) Join
2) Aggregation (count, max, and min).
3) Select where query conditions contain (one, two, and 3 clauses)

Table 3 illustrates how our model dramatically drops the average of retrieved encrypted candidate records for a query. That means we successfully enable the cloud provider to process queries over encrypted data without decrypting them. We can notice that our model is more efficient when the query condition involves more clauses and the operation between the clauses is the AND. If the OR operation is present in the query condition, our model returns more candidate encrypted tuples because of the nature of the OR operation. Moreover, if the condition contains sensitive columns that are ranges, our model will experience more candidate encrypted tuples to decrypt at the QM. That will add more delay because of the decryption processes and filtering out unrelated tuples from the result.

**Table 3.** Average number of retrieved encrypted candidate tuples for different select query operations and clauses. N.R denotes the number of records in a table.

| N.R | Average number of retrieved encrypted candidate tuples from Cloud | | | | |
|---|---|---|---|---|---|
| | *1 clause* | *2 clauses (AND)* | *3 clauses (AND)* | *2 clauses (OR)* | *3 clauses (OR)* |
| 10K | 2035 | 303 | 44 | 3408 | 6227 |
| 20K | 4758 | 583 | 96 | 7281 | 12106 |

In Fig. 3, we compare the delay of our model and the delay of traditional database system TDBS (where data are not encrypted). We ran select queries where condition contains three clauses but with different operations, AND and OR. We can see that the delays when the AND operation is present in the condition is significantly small when compared with the delay when the OR operation is present in the condition.

Fig. 4 shows the percent of encrypted records reprocessed at the QM after the decryption in union and intersection operations. In the intersection, we found that our model filters out approximately 83 percent of uncommon encrypted tuples at the cloud, and only about 17 percent of the entire result was common. This is because we can't eliminate the duplication at the cloud when we have range attributes, e.g. salary. Similarly, in *join* and *union*, less than 30 percent of the results were filtered out at the QM.

According to this, we can say that our model efficiently reduces the delay that results from unnecessary decryption processes.

In fig. 5, we can see that the execution time is neglectable when the aggregation query is based on sensitive columns that are not ranges. In such case, the QM doesn't perform decryption; however, the QM will execute the aggregation query over decrypted data if the query is based on a range column, that is why we see the execution time is higher than the first case (non-range columns).

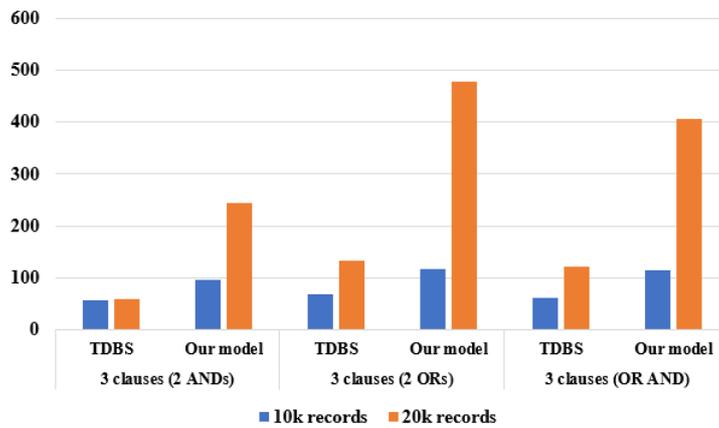

**Fig. 3.** Delays in our model and how they are affected by the number of clauses and operations in a query condition.

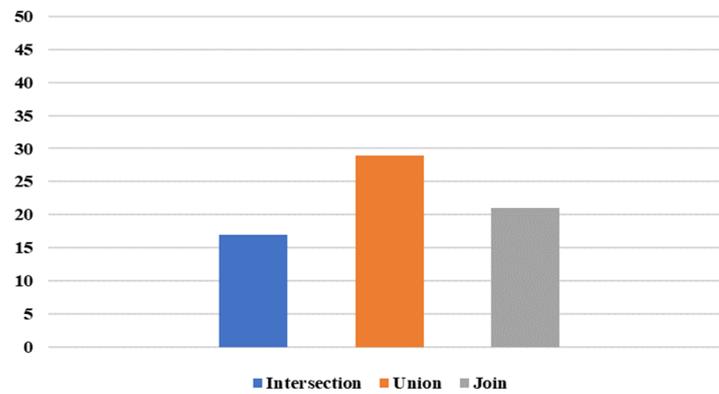

**Fig. 4.** The % of tuples reprocessed at the QM after decrypting the intersection, union, and join results.

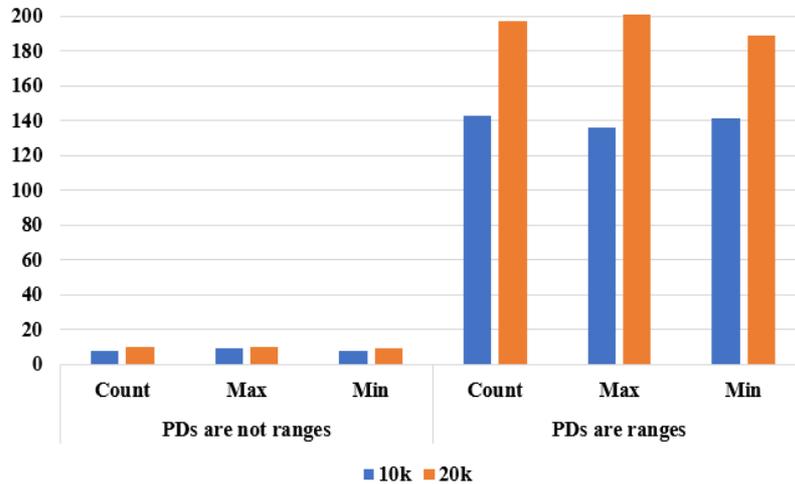

**Fig. 5.** The execution time in *ms* for aggregation operators (count, max, and min) when the query condition has sensitive columns that are not ranges vs. sensitive columns that are ranges.

## 5    Conclusion

Cloud computing is an attractive computing environment for all types of users and companies. However, privacy breaches, not only by malicious attackers but also by curious providers, is the downside of such service, because users lose access control over outsourced data. Data encryption is an effective solution for this problem. However, searching over encrypted data is challenging, especially if a randomized encryption algorithm is used for the encryption. In this research, we design a secure model to process queries and retrieve encrypted tuples from encrypted databases that preserve privacy and is efficient at the same time. The model is based on classifying columns as sensitive and non-sensitive, where only sensitive columns are encrypted. Furthermore, the table's owner predefines the possible PDs and ranges for each sensitive column. Then for each table, we make a DS of the PDs, and create a bit vector whose length is equal to the number of PDs. We then map each PD to a specific bit where this bit is set to one if and only if the value of the corresponding column is equal or belongs to this PD. We use the bits to retrieve candidate tuples for a certain query that minimize the range of the retrieved encrypted tuples. The encrypted table in the cloud will have an additional column (the reference column) to store bit vectors. To make the model practical, we facilitate it to support most relational algebra operators. We performed different experiments to test the efficiency of the model and found that it reduces the range of retrieved encrypted tuples to about 30 percent of the whole encrypted tuples in a table.

## References

1. Almakdi, Sultan, Panda Brajendra. "Secure and Efficient Query Processing Technique for Encrypted Databases in Cloud." 2019 2nd International Conference on Data Intelligence and Security (ICDIS). IEEE, 2019.
2. Alsirhani, Amjad, Peter Bodorik, and Srinivas Sampalli. "Improving database security in cloud computing by fragmentation of data." Computer and Applications (ICCA), 2017 International Conference on. IEEE, 2017.
3. Cui, S., Asghar, M. R., Galbraith, S. D., & Russello, G. (2017, June). P-McDb: Privacy-preserving search using multi-cloud encrypted databases. In 2017 IEEE 10th International Conference on Cloud Computing (CLOUD) (pp. 334-341). IEEE.
4. D. Cash, J. Jaeger, S. Jarecki, C. S. Jutla, H. Krawczyk, M. Rosu, and M. Steiner, "Dynamic searchable encryption in very-large databases: Data structures and implementation," in NDSS 2014, The Internet Society, 2014.
5. D. Cash, P. Grubbs, J. Perry, and T. Ristenpart, "Leakage-abuse attacks against searchable encryption," in SIGSAC 2015 (I. Ray, N. Li, and C. Kruegel, eds.), pp. 668–679, ACM, 2015.
6. E. Stefanov, C. Papamanthou, and E. Shi, "Practical dynamic searchable encryption with small leakage.," in NDSS 2013, vol. 71, pp. 72–75, 2014.
7. Gentry, C. (2009, May). Fully homomorphic encryption using ideal lattices. In Stoc (Vol. 9, No. 2009, pp. 169-178).
8. Gentry, C., Halevi, S., & Smart, N. P. (2012, April). Fully homomorphic encryption with polylog overhead. In Annual International Conference on the Theory and Applications of Cryptographic Techniques (pp. 465-482). Springer, Berlin, Heidelberg.
9. Hacıgümüş, Hakan, Bala Iyer, Chen Li, and Sharad Mehrotra. "Executing SQL over encrypted data in the database-service-provider model." In Proceedings of the 2002 ACM SIGMOD international conference on Management of data, pp. 216-227. ACM, 2002.
10. Hacigumus, Vahit Hakan, Balakrishna Raghavendra Iyer, and Sharad Mehrotra. "Query optimization in encrypted database systems." U.S. Patent No. 7,685,437. 23 Mar. 2010.
11. Hore, Bijit, et al. "Secure multidimensional range queries over outsourced data." The VLDB Journal 21.3 (2012): 333-358.
12. Hore, Bijit, Sharad Mehrotra, and Gene Tsudik. "A privacy-preserving index for range queries." Proceedings of the Thirtieth international conference on Very large data bases-Volume 30. VLDB Endowment, 2004.
13. K. Li, W. Zhang, C. Yang, and N. Yu, "Security Analysis on One-to-Many Order Preserving Encryption-Based Cloud Data Search," IEEE Transactions on Information Forensics and Security,vol.10,no.9,pp.1918–1926,2015.
14. Kamara, Seny, and Tarik Moataz. "SQL on structurally-encrypted databases." In International Conference on the Theory and Application of Cryptology and Information Security, pp. 149-180. Springer, Cham, 2018.
15. Li, Jin, Zheli Liu, Xiaofeng Chen, Fatos Xhafa, Xiao Tan, and Duncan S. Wong. "L-EncDB: A lightweight framework for privacy-preserving data queries in cloud computing." Knowledge-Based Systems 79 (2015): 18-26.
16. Liu, D., & Wang, S. (2013). Nonlinear order preserving index for encrypted database query in service cloud environments. Concurrency and Computation: Practice and Experience, 25(13), 1967-1984.
17. Liu, G., Yang, G., Wang, H., Xiang, Y., & Dai, H. (2018). A Novel Secure Scheme for Supporting Complex SQL Queries over Encrypted Databases in Cloud Computing. Security and Communication Networks, 2018.

18. Liu, X., Choo, K. K. R., Deng, R. H., Lu, R., & Weng, J. (2018). Efficient and privacy-preserving outsourced calculation of rational numbers. IEEE Transactions on Dependable and Secure Computing, 15(1), 27-39.
19. Liu, Z., Chen, X., Yang, J., Jia, C., & You, I. (2016). New order preserving encryption model for outsourced databases in cloud environments. Journal of Network and Computer Applications, 59, 198-207.
20. M. R. Asghar, G. Russello, B. Crispo, and M. Ion, "Supporting complex queries and access policies for multi-user encrypted databases," in CCSW 2013 (A. Juels and B. Parno, eds.), pp. 77–88, ACM, 2013.
21. Omran, Osama M., "Data Partitioning Methods to Process Queries on Encrypted Databases on the Cloud" (2016).Theses and Dissertations. 1580.
22. Poddar, R., Boelter, T., & Popa, R. A. (2016). Arx: A strongly encrypted database system. IACR Cryptology ePrint Archive, 2016, 591.
23. Popa, Raluca Ada, et al. "CryptDB: processing queries on an encrypted database." Communications of the ACM 55.9 (2012): 103-111.
24. R. Agrawal, J. Kiernan, R. Srikant, and Y. R. Xu, "Order preserving encryption for numeric data," in Proceedings of the ACM SIGMOD International Conference on Management of Data (SIGMOD '04), pp. 563–574, ACM, Paris, France, June 2004.
25. Raybourn, Tracey, Jong Kwan Lee, and Ray Kresman. "On Privacy Preserving Encrypted Data Stores." Multimedia and Ubiquitous Engineering. Springer, Dordrecht, 2013. 219-226.
26. Raybourn, Tracey. Bucketization Techniques for Encrypted Databases: Quantifying the Impact of Query Distributions. Diss. Bowling Green State University, 2013.
27. Shastri, Samraddhi, Ray Kresman, and Jong Kwan Lee. "An Improved Algorithm for Querying Encrypted Data in the Cloud." Communication Systems and Network Technologies (CSNT), 2015 Fifth International Conference on. IEEE, 2015.
28. Tu, Stephen, et al. "Processing analytical queries over encrypted data." Proceedings of the VLDB Endowment. Vol. 6. No. 5. VLDB Endowment, 2013.
29. Wang, Jieping, et al. "Bucket-based authentication for outsourced databases." Concurrency and Computation: Practice and Experience 22.9 (2010): 1160-1180.
30. Wang, W., Hu, Y., Chen, L., Huang, X., & Sunar, B. (2015). Exploring the feasibility of fully homomorphic encryption. IEEE Transactions on Computers, 64(3), 698-706.
31. Wong, Wai Kit, et al. "Secure query processing with data interoperability in a cloud database environment." Proceedings of the 2014 ACM SIGMOD international conference on Management of data. ACM, 2014.
32. Y. Zhang, J. Katz, and C. Papamanthou, "All your queries are belong to us: The power of file-injection attacks on searchable encryption," in USENIX Security 2016, pp. 707–720, USENIX Association, 2016.
33. Harkins, Dan. "Synthetic initialization vector (siv) authenticated encryption using the advanced encryption standard (aes)." (2008).